\documentclass{ws-ijmpa}
\usepackage[super,compress]{cite}
\usepackage{graphicx}
\begin{document}
\markboth{Ronan McNulty}
{Results and Future Plans on
Central Exclusive Production with the LHCb Detector}

%
\catchline{}{}{}{}{}
%

\title
{RESULTS AND FUTURE PLANS ON CENTRAL EXCLUSIVE PRODUCTION WITH 
THE LHCb DETECTOR}

\author{RONAN MCNULTY}

\address{School of Physics,
University College Dublin, \\
Dublin 4, Ireland\\
ronan.mcnulty@ucd.ie}

\maketitle


\begin{abstract}
The LHCb detector and LHC running conditions
are ideally suited to measuring central exclusive production.
Several recent measurements of exclusive dimuon, single and double charmonia are
reviewed.
The potential for future measurements across a broad range of physics channels is discussed.

\end{abstract}

\ccode{PACS numbers:}


\section{Introduction}

Proton-proton collisions at LHC energies  
usually produce hundreds of charged and neutral particles.
However, when 
colourless propagators are involved and, in addition, the protons remain intact,
the final state consists only of protons and
that produced by the fusion of the propagators.
Experimentally,
this leads to a unique signature for central exclusive production (CEP)~\cite{albrow_review,kmr_review}
of  a small number of particles in the central region, either those 
produced directly or their decay products, and two rapidity gaps that extend to 
the outgoing protons in the far-forward direction that are minimally deflected from their
initial trajectories.
Although designed with b-physics in mind, the LHCb detector is well suited to the
detection and study of CEP due to its 
ability to trigger and reconstruct low mass central systems, 
its good particle identification, its large pseudorapidity
acceptance, and the running conditions of the LHC.

This short review of CEP at LHCb summarises the measurements that
have been performed and looks ahead to what can be expected 
in the future.  
A brief description of the LHCb detector and the features that make it
suitable for identifying CEP is given in Sec.~\ref{sec:det}.
The experience of the last four years of data-taking and analysis
has highlighted a difficulty in
relating the experimental observations to the theoretical predictions for
elastic CEP, and this issue and the approach taken to overcome it,
is described in Sec.~\ref{sec:exc}.
Following this, the measurements are presented divided up by the
production mechanism: 
photon-pomeron fusion is dealt with in Sec.~\ref{sec:photpom}; 
two photon physics is described in Sec.~\ref{sec:photphot};
and double pomeron exchange (DPE) is discussed in Sec.~\ref{sec:dpe}.
In each section, the current measurements are reviewed before prospects
for the future are outlined.
Sec.~\ref{sec:summary} presents a summary and a forward look to a new
sub-detector 
being installed for the upcoming LHC running,
which will significantly mitigate the problems discussed in Sec.~\ref{sec:exc}.

\section{Suitability of the LHCb Detector for Measuring CEP}
\label{sec:det}
The LHCb detector~\cite{lhcb} 
is fully instrumented between pseudorapidities, $\eta$, of 2 and 4.5 and 
includes a high-precision tracking system
consisting of a silicon-strip vertex detector (VELO) surrounding the $pp$
interaction region~\cite{velo}, a large-area silicon-strip detector located
upstream of a dipole magnet with a bending power of about
$4{\rm\,Tm}$, and three stations of silicon-strip detectors and straw
drift tubes~\cite{ot} placed downstream of the magnet.
Different types of charged hadrons are distinguished using information
from two ring-imaging Cherenkov detectors~\cite{rich}. Photon, electron and
hadron candidates are identified by a calorimeter system consisting of
scintillating-pad (SPD) and preshower detectors, an electromagnetic
calorimeter and a hadronic calorimeter. Muons are identified by a
system composed of alternating layers of iron and multiwire
proportional chambers~\cite{muon}.
LHCb accumulated an integrated luminosity of
$37 {\rm pb}^{-1}$ in 2010 at a collider centre-of-mass energy, $\sqrt{s}=7$ TeV,
$1 {\rm fb}^{-1}$ in 2011 at $\sqrt{s}=7$ TeV,
and $2 {\rm fb}^{-1}$ in 2012 at $\sqrt{s}=8$ TeV.

LHCb is
often described as a {\it forward} spectrometer.  However, 
in the context of diffractive physics,
where for an object of mass, $m$, 
the pseudorapidity lies between $\pm \log(\sqrt{s}/m)$, 
there is a large acceptance for
{\it central} exclusive production.
The determination of the {\it exclusivity} of an event depends on no activity being seen
in an active detection region that extends over as large a pseudorapidity range as possible.  
The VELO detector consists of 21 planes of silicon
microstrip detectors placed normal to the beamline and around the interaction point.
Charged particles passing through it are recorded with over 99\% efficiency,
while noise rates are very low, typically below five hits out of 86,000 strips.
At least three planes need to be traversed before a track can be reconstructed from the 
charge deposits. 
Because the interaction point is inside the VELO acceptance, both
forward and backward going tracks can be reconstructed, 
and this extends the sensitivity of LHCb to charged
particles to $-3.5<\eta<-1.5$ and $1.5<\eta<5$.  Although it is not possible to measure
the momentum of particles throughout this region, a charged particle veto corresponding
to about 5.5 units of pseudorapidity can be obtained.

The triggering capability of the LHCb detector, being designed for low mass objects, 
is well suited to CEP.  It consists of a two-stage system, a fast hardware trigger followed by
a software trigger that applies a full event reconstruction.
For CEP,
the hardware stage triggers on muons with transverse momentum above \mbox{400 MeV}, 
or electromagnetic or hadronic transverse energy  above 1000 MeV, 
all of which are placed in coincidence with a charged
multiplicity of less than 10 deposits in the SPD.
The software trigger is configured to select a variety of final states, including those with
two muons, two electrons, two photons, two pions, two kaons,
or to explicitly select candidates for $\phi$ or $D$ mesons through their decays to 
pions and kaons.  

The data-taking conditions at LHCb are advantageous for the selection of CEP events.
The detection of an exclusive event usually requires ensuring no activity in the detector
apart from the central system.
However, in order to achieve high luminosities, the density of protons in the LHC
bunches is high, leading to several proton-proton interactions every beam-crossing.  
A potential exclusive signal from  one of these is obscured by the activity from the others.
Unlike ATLAS and CMS, where there were typically 20 interactions
per beam-crossing in the 2012 data-taking,
the beams are defocused at LHCb, resulting in an average of about 1.5 proton-proton
interactions per beam-crossing.  Consequently, about 20\% of the total luminosity has a
single interaction and is usable for CEP.

\section{Measuring Exclusivity with LHCb}
\label{sec:exc}

It is worth spending a little time to discuss exactly what LHCb can measure.
Direct detection of particles is limited to the pseudorapidity region in which the detector is
instrumented.  Any activity outside that region can not be directly seen, although it may be
indirectly inferred, in which case the limiting uncertainty becomes the degree
of trust in the underlying assumptions.
Theoretical calculations for CEP are generally made for the elastic process in which 
the central system is separated from the protons by two rapidity gaps.
This ideal situation can be spoiled by two effects: additional gluon radiations between
the two protons or between the protons and the central system; or proton dissociation where
the pomeron or photon perturbs the proton sufficiently for it to break up.
The theoretical modelling of these processes is difficult: see Ref.\,[\citen{kmrpom,dime}] for
further discussion.  Both effects lead to additional particles that reduce the rapidity gap; however,
these are often produced in the very forward direction and may not be experimentally detected.

Strictly speaking, LHCb does not measure CEP because it does not have $4\pi$ coverage
and can not guarantee the total exclusivity of the event.
In order to compare with theoretical calculation of CEP, there are two alternatives.
LHCb could simply report measurements that are exclusive {\it inside the LHCb acceptance}.  However, this presents difficulties for theorists wishing to use the results as 
firstly, they would need an accurate model of the detector efficiency, acceptance and LHC machine 
parameters and secondly, they would need to model the inelastic backgrounds.
The alternative is to make model dependent assumptions to estimate the fraction of 
the exclusively observed signal which is due to the elastic CEP component.
This requires an indirect measurement of the activity outside the active region.
The best variable for this purpose appears to be the transverse momentum, $p_T$, 
of the central system.

Under the assumption that the more the proton is perturbed, the more chance it has of
breaking up, there is a correlation between proton dissociation and  the $p_T$
of the central system.  In contrast, for elastic CEP events, the transverse momentum of 
the central system is generally small.
The approach taken in the LHCb analyses to date is to invoke
Regge theory to describe the shape of the transverse momentum distribution.
Regge theory
assumes the relationship $d\sigma/dt \propto \exp(b^\prime t)$ for a wide class of diffractive events, 
where $b^\prime$ is a constant for a given process, 
$t\approx -p_{T(p)}^2$ is the four-momentum transfer squared at one of the proton-pomeron vertices,
and $p_{T(p)}$ is the transverse momentum of the outgoing proton.
Consequently, the $p_T^2$ distribution 
 for the central system can be described by the
functional form $\exp(-bp_T^2)$.
The CEP signal is found to have a large $b$ value  at LHC energies, typically 
$4 {\rm \ GeV}^{-2}$ for DPE processes, 
$6 {\rm \ GeV}^{-2}$ for photon-pomeron fusion,
and above $10 {\rm\ GeV}^{-2}$ for two-photon physics processes.
The inelastic processes have fitted $b$ values of 
$1 {\rm \ GeV}^{-2}$ or smaller. 
Fitting the $p_T^2$ spectrum in data with two components, one for 
elastic CEP and one for the inelastic background,
thus provides a mechanism to extract the signal fraction:
the only assumption is that each can be modelled by an exponential function.
This was tested by the H1 collaboration~\cite{h1} in the case of $J/\psi$ photoproduction:
an exponential was found to describe the
elastic data while a power law, that approximates to an exponential
at small $p_T^2$, was found to fit the proton dissociative component.

\begin{figure}[b]
\centerline{\includegraphics[height=5cm]{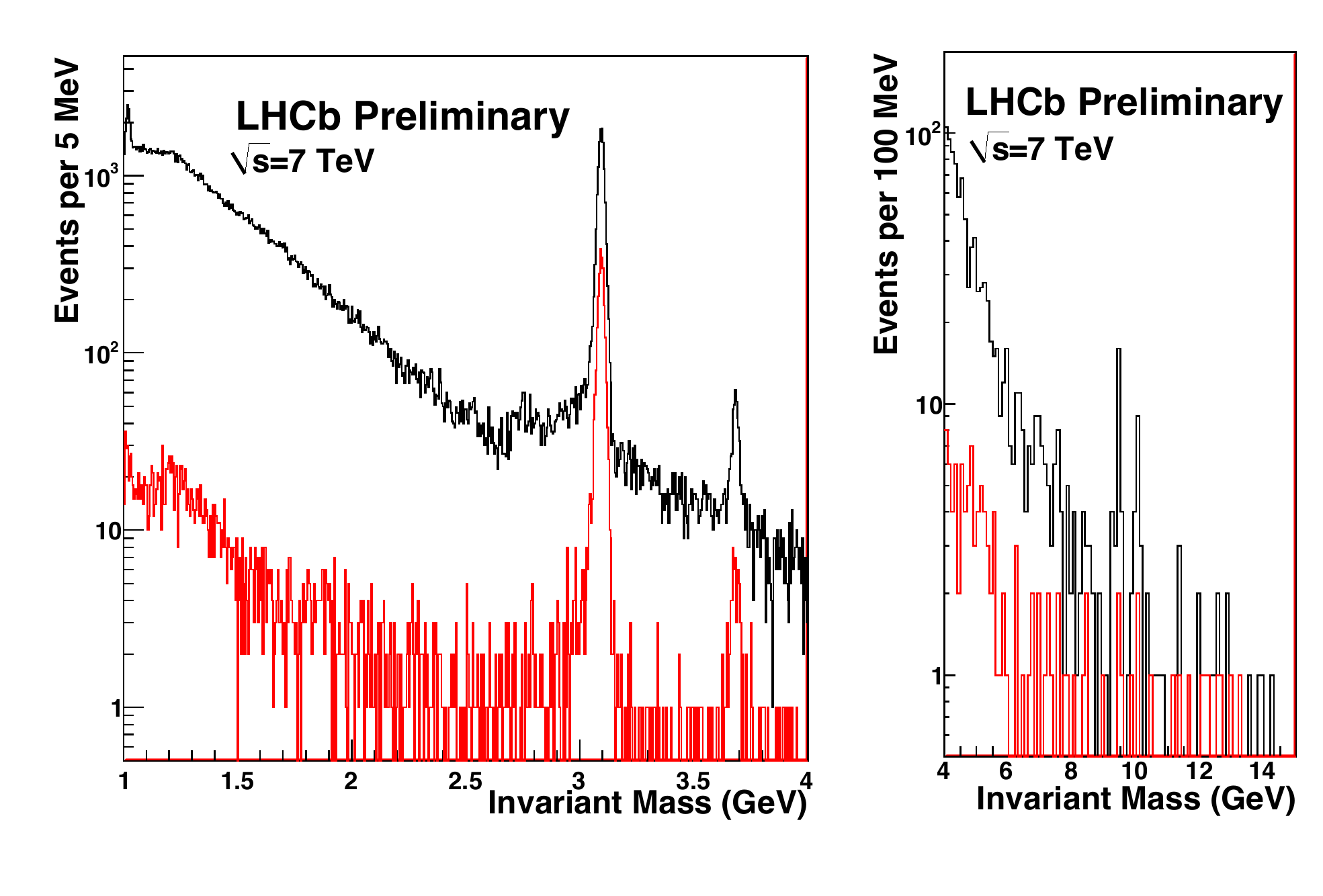}}
\caption{
Invariant mass of dimuons in $37 {\rm\ pb}^{-1}$ of data
after the low-multiplicity dimuon trigger (upper histogram, black)
and after requiring no other activity inside LHCb (lower histogram, red).
The discontinuity at 2.5 GeV is due to a trigger threshold.
\label{fig:mu2mass}}
\end{figure}

\section{Photon-Pomeron Fusion}
\label{sec:photpom}
\begin{figure}[b]
\centerline{\includegraphics[height=5cm]{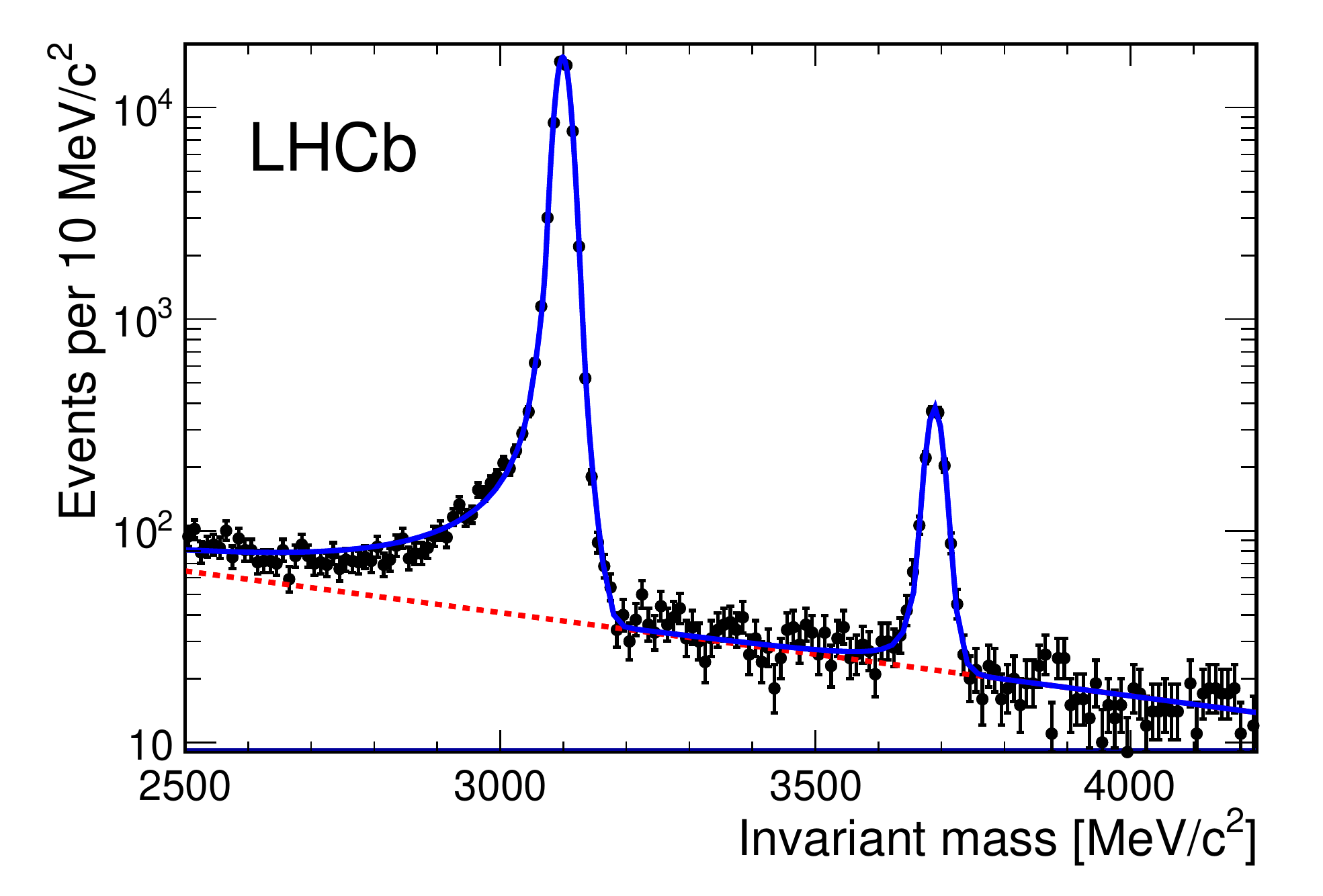}}
\caption{
Invariant mass of dimuon candidates in $930 {\rm\ pb}^{-1}$ of data.
\label{fig:jmass}}
\end{figure}

The LHCb collaboration has made two measurements of $J/\psi$ and $\psi(2S)$
production at $\sqrt{s}=7$ TeV, 
one with an integrated luminosity of 
$37{\rm \ pb}^{-1}$ (2010 data)~\cite{lhcbj2010}, and one
with $930{\rm \ pb}^{-1}$ (2011 data)~\cite{lhcbj2011}.
The selection of events starts by triggering on low multiplicity events
containing two muons.  The events are then selected as exclusive inside the
LHCb acceptance by requiring no additional charged tracks or neutral deposits in 
the detector.
The invariant mass of the two muons after the trigger and after the selection
is shown in Fig.~\ref{fig:mu2mass} for the $37{\rm \ pb}^{-1}$ sample.
Within a falling continuum, there are clear signals after the trigger requirements
for several vector mesons: $\phi,J/\psi,\psi(2S),\Upsilon(1S),\Upsilon(2S)$.
With the additional exclusivity requirements, only charmonia signals remain 
visible in this limited data sample.
Candidate events for $J/\psi$ and $\psi(2S)$ mesons in the larger $930{\rm \ pb}^{-1}$
sample can be seen in Fig.~\ref{fig:jmass}.  No background subtraction has been performed
in these plots: the $J/\psi$ is over two order of magnitude above the continuum while the
$\psi(2S)$ signal is somewhat less prominent, mainly due to its lower branching fraction
to two muons.

Three backgrounds are considered in extracting the elastic CEP signal: 
non-resonant dimuon production; feed-down from other mesons; 
and inelastic $J/\psi$ production.
The non-resonant background
is determined from the fit shown in Fig.~\ref{fig:jmass}.
Feed-down is only considered for the $J/\psi$ selection and can come from 
$\chi_{c0},\chi_{c1},\chi_{c2}$ or $\psi(2S)$ decays,
with the other decay products being below the threshold for detection or outside
the LHCb acceptance.  Feed-down from $\chi_c\rightarrow J/\psi\gamma$
is evaluated to be $(7.6\pm0.9)\%$ by selecting events in which the photon is seen and using the simulation
to estimate the number of events in which it would be undetected.  Feed-down from the
decays $\psi(2S)\rightarrow J/\psi X$
is estimated from the simulation, which has been normalised to the
observed number of events from the decay $\psi(2S)\rightarrow\mu\mu$,
and contributes $(2.5\pm 0.2)\%$ of the $J/\psi$ sample.

\begin{figure}[b]
{\includegraphics[width=0.49\linewidth]{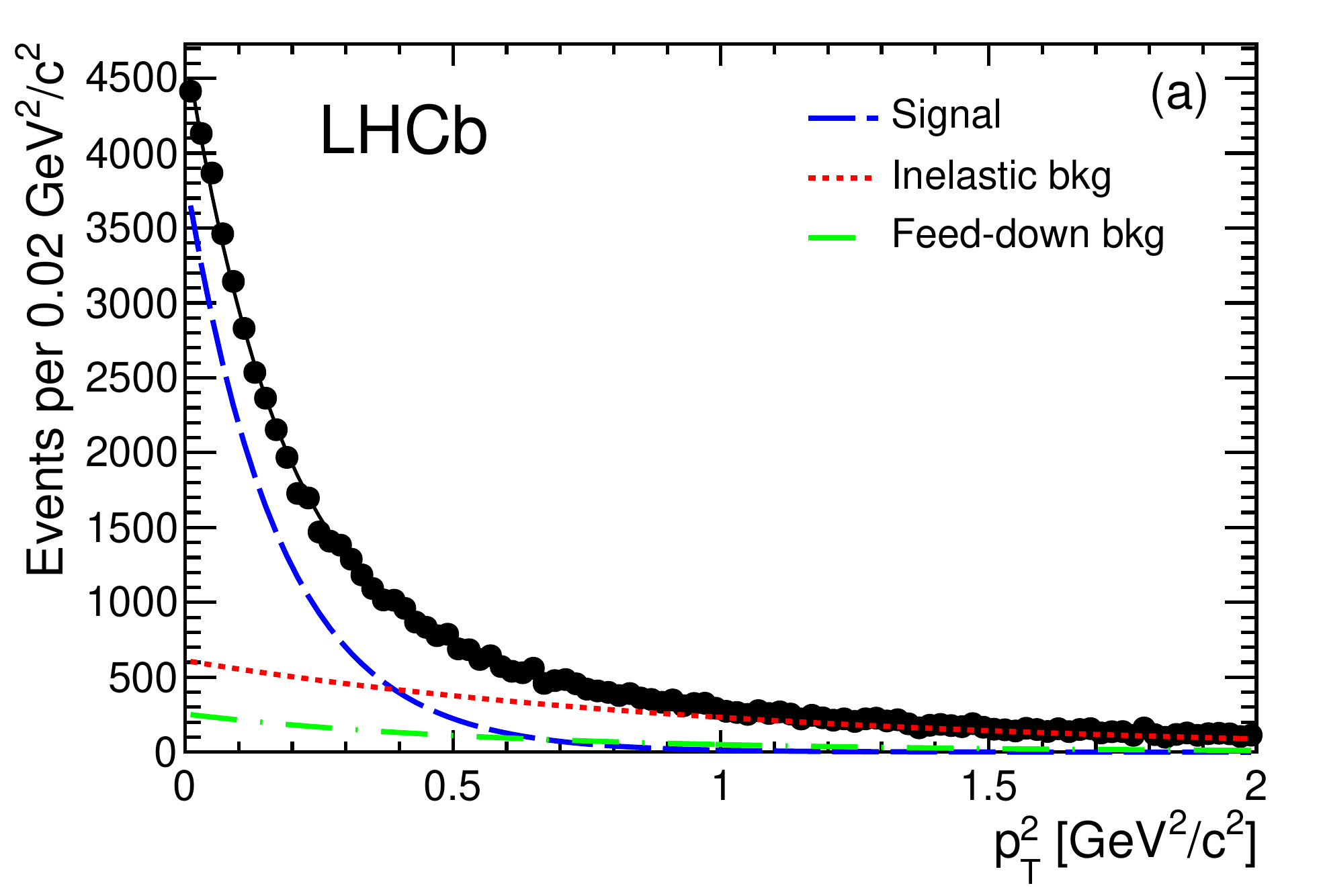}}
{\includegraphics[width=0.49\linewidth]{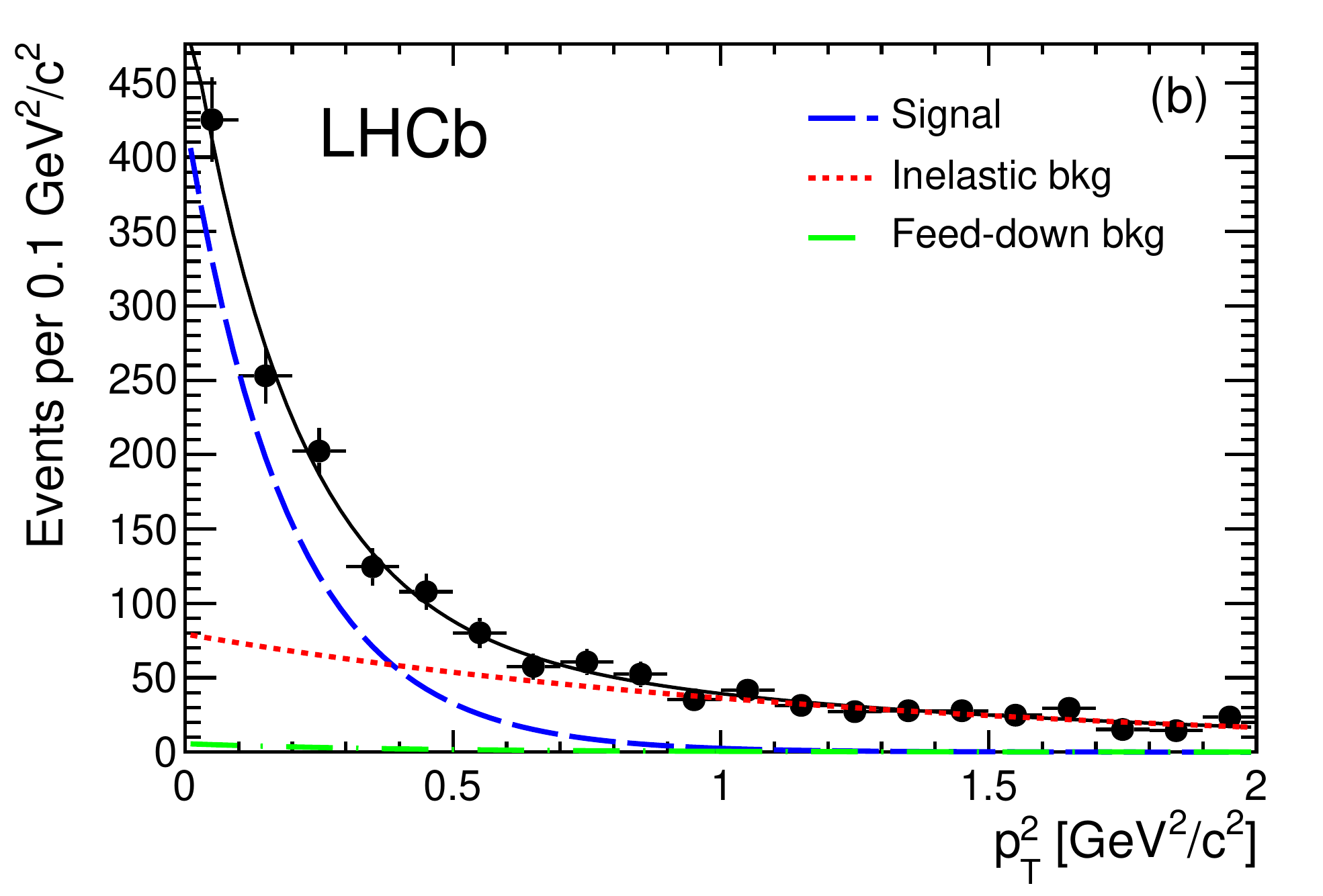}}
\caption{
Transverse momentum squared of (a) $J/\psi$ and (b) $\psi(2S)$ candidates.
The fitted contributions from the CEP signal, the inelastic and feed-down backgrounds
are indicated in the legend.
\label{fig:pt2}}
\end{figure}

The third background source is  
the largest and is also the most poorly determined for this analysis and all other CEP analyses
that LHCb have performed or are likely to perform with the data taken to date.
As discussed in Sec.~\ref{sec:exc}, this consists of 
centrally produced $J/\psi$ or $\psi(2S)$ mesons which appear exclusive inside
the LHCb acceptance, but have activity outside of the active area of the detector, originating either
from additional gluon radiations or low mass diffractive dissociation of one or both protons.
Assuming that the $p_T^2$ distribution for both the elastic and inelastic components can
be described by exponential functions, $\exp(-bp_T^2)$, a fit to the data is performed to
determine the $b$ values and the relative sizes of both components.
The results are shown in Fig.~\ref{fig:pt2} and an overall purity of $0.592\pm 0.012$ is obtained
for the $J/\psi$ sample and $0.52\pm 0.07$ for the $\psi(2S)$ sample.
It is also worth noting that the fitted $b$ values  are consistent with the photoproduction results
from the H1 collaboration~\cite{h1}, once the difference in the centre-of-mass of 
the photon-proton system has been taken into account.

\begin{figure}[b]
{\includegraphics[width=0.49\linewidth]{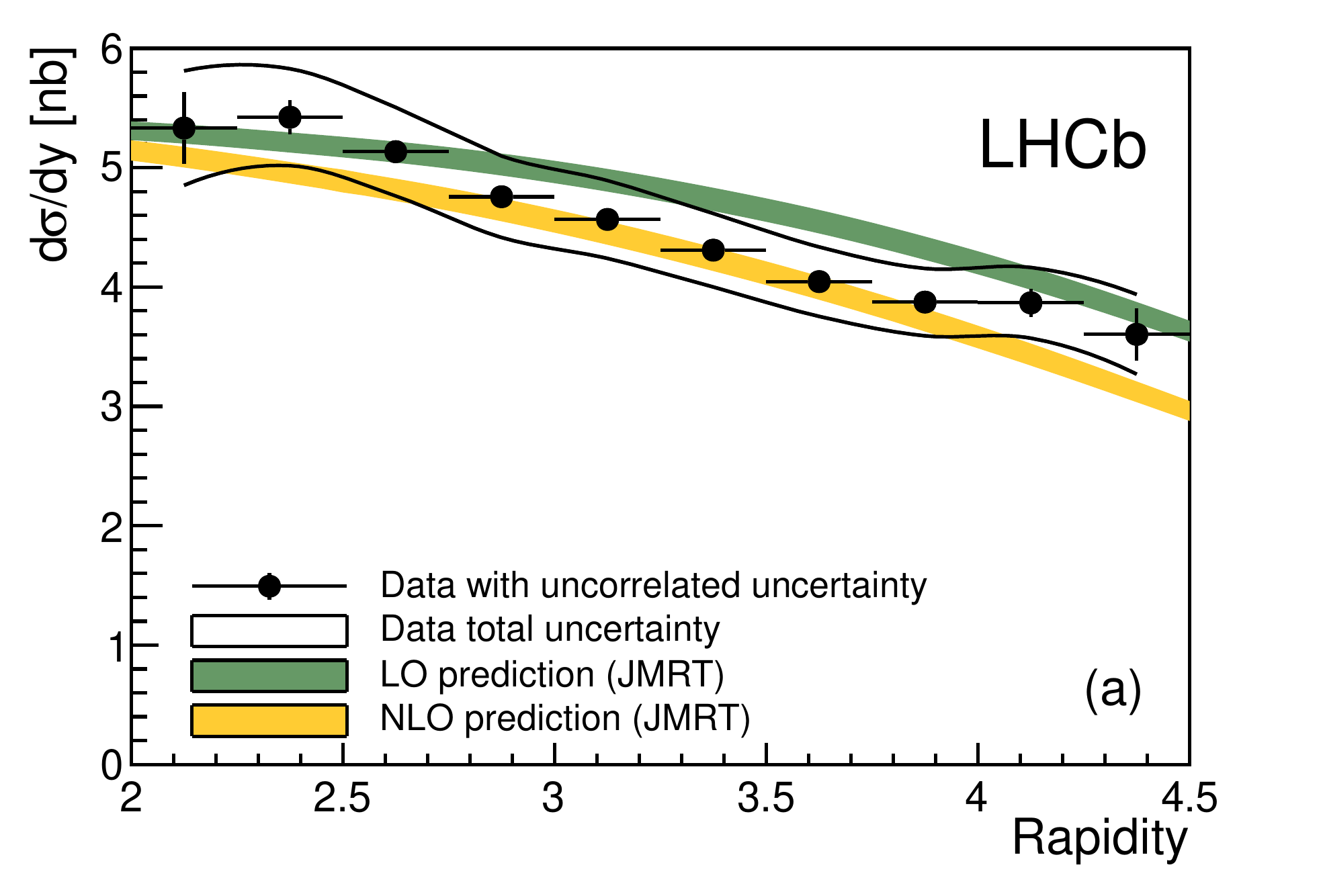}}
{\includegraphics[width=0.49\linewidth]{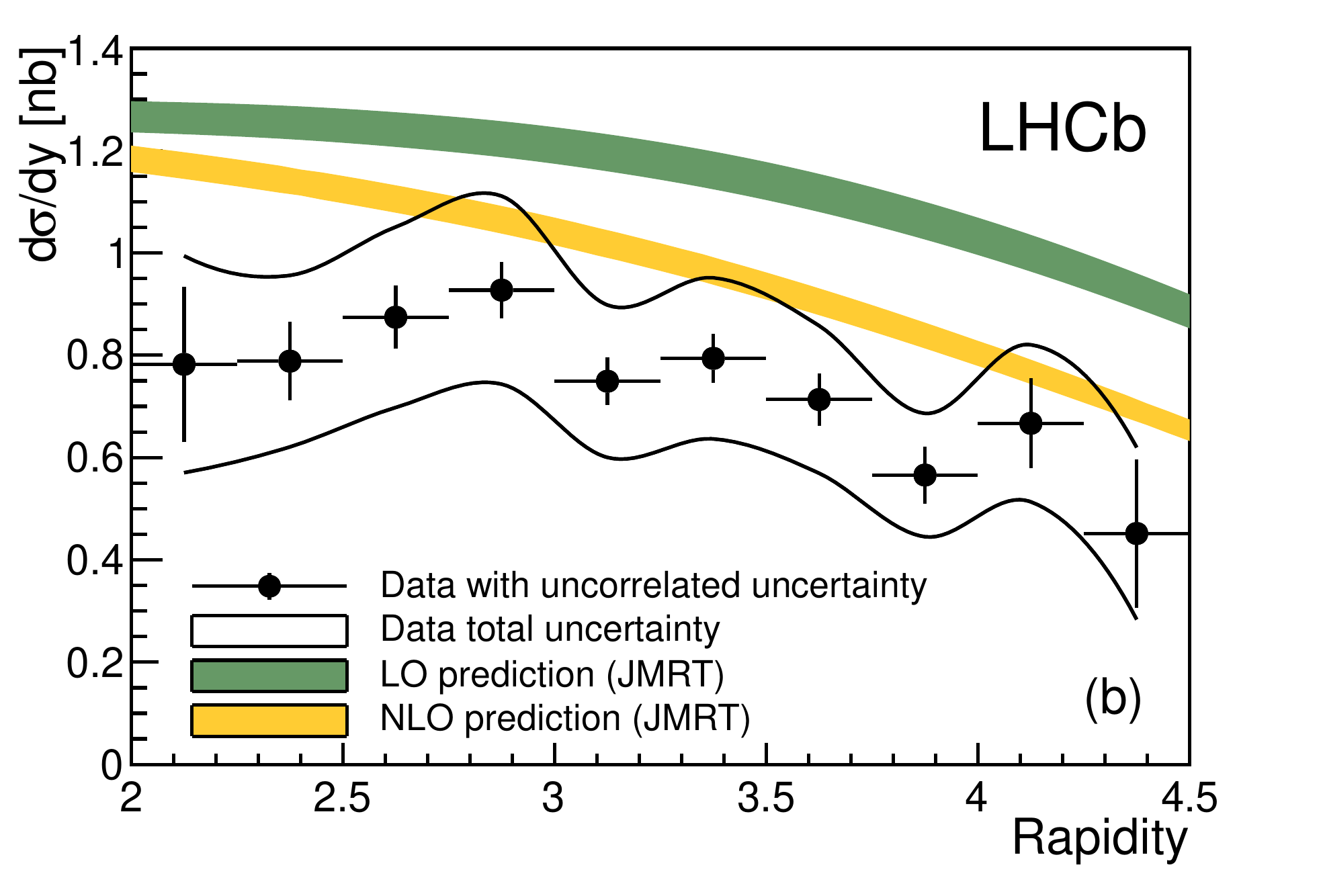}}
\caption{
Differential cross-section for 
(a) $J/\psi$ and (b) $\psi(2S)$ 
compared to LO and NLO predictions.
\label{fig:cs}}
\end{figure}

After correcting for the detector efficiency and acceptance, 
the differential cross-section as a function of rapidity, $y$, is calculated and is shown
in Fig.~\ref{fig:cs} compared to predictions at LO and `NLO' from Ref.\,[\citen{jmrt1,jmrt2}].  
The `NLO' calculation is not a
full next-to-leading-order calculation but includes the dominant effects.
The experimental points are plotted with their statistical and total uncertainties.  
Most of the total uncertainty is correlated bin-to-bin and so the overall shape is well determined
by the data, which fits the `NLO' predictions better than LO, 
for both the $J/\psi$ and $\psi(2S)$ mesons.

\subsection{Future analyses}

Fig.~\ref{fig:mu2mass} shows clear signals for other vector mesons after the trigger requirements
with just $37 {\rm\ pb}^{-1}$ of data.  With a total of about $3 {\rm\ fb}^{-1}$ of data taken
at $\sqrt{s}=7$ and 8 TeV, large samples of $\phi$ and $\Upsilon$ produced
in a low multiplicity environment have been collected.
Analyses are ongoing to determine how much of each of these is elastic CEP.

For the CEP $J/\psi$ analysis, the forthcoming LHC running at $\sqrt{s}=13$ TeV presents a
unique opportunity to constrain the gluon 
parton distribution function (PDF) at very low fractional proton energies, $x$,
down to $2\times 10^{-6}$, where it is essentially unknown.
As the gluon PDF enters squared in the prediction of the $J/\psi$ cross-section~\cite{ryskin}, 
the current results at $\sqrt{s}=7$ TeV can, in principle, be used to constrain it.  
However, there are large scale uncertainties in the theory which mean that, to date,
the results on CEP of vector mesons are not currently included in the 
global PDF fits.~\cite{mstw,cteq,nnpdf}
However, by taking the 
ratio of cross-sections at different centre-of-mass energies, the scale uncertainties
should be significantly reduced~\cite{rojo}, 
leading to a more precise measurement of the ratio of PDFs
which can be used to constrain the gluon at low-$x$ with a much reduced theoretical
uncertainty.

\section{Two-Photon Physics}
\label{sec:photphot}

Physics from the fusion of two photons in proton-proton collisions
can be accessed when the central system
can not be produced by the strong force.  
The non-resonant elastic production of lepton pairs
is a QED process that can be predicted with about 1\% precision~\cite{serbo}, 
and so the experimental
measurement of this process is, in principle, an excellent way to determine the machine
luminosity precisely.

\begin{figure}[b]
{\includegraphics[width=0.49\linewidth]{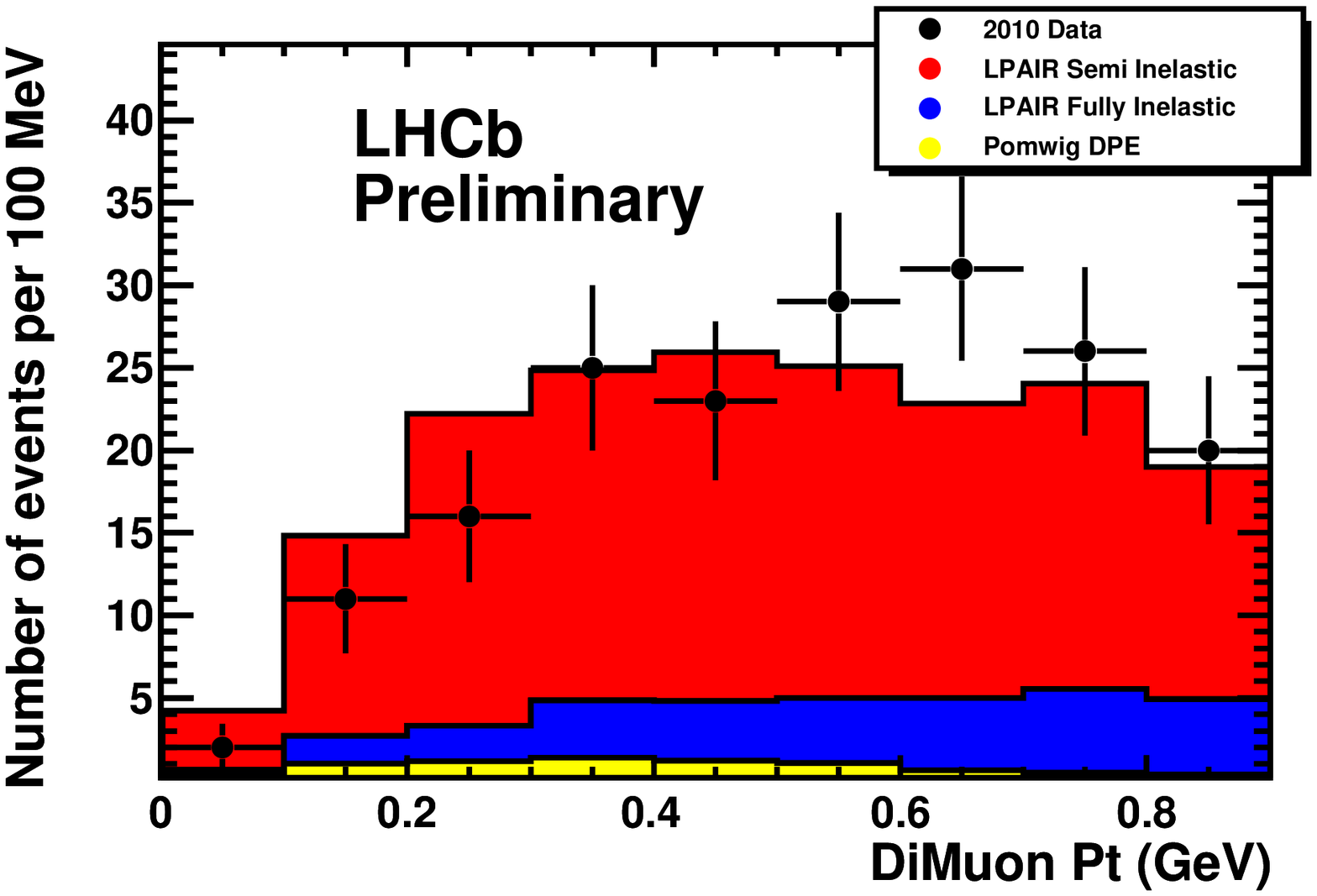}}
{\includegraphics[width=0.49\linewidth]{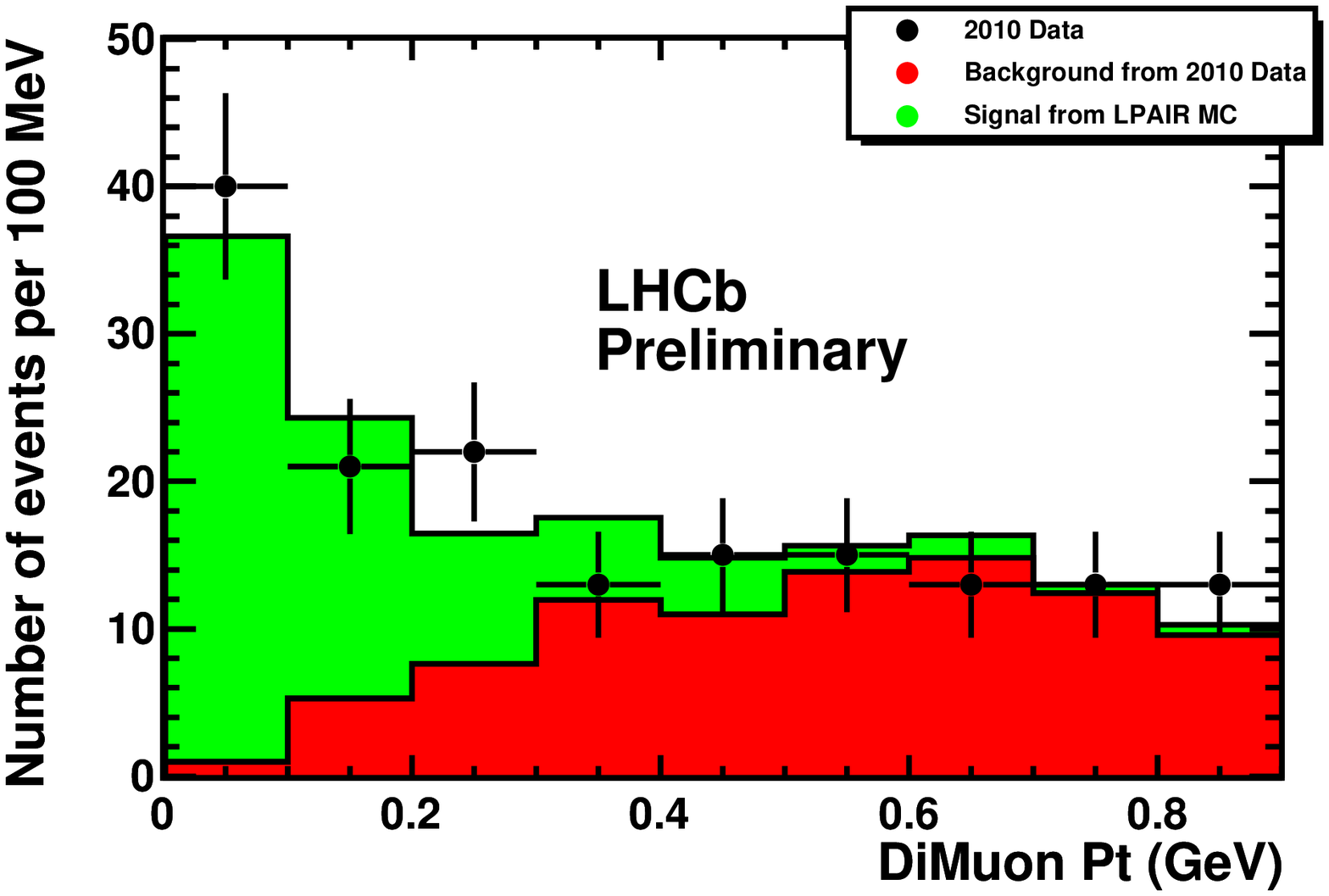}}
\caption{
Transverse momentum of dimuons that have an invariant mass above 2.5 GeV and are not 
consistent with vector meson production.  The plot on the left shows events with more
than two tracks compared to expectations for inelastic dimuon production.  The plot on
the right shows events with exactly two tracks and no other activity inside the LHCb detector.
The shape of the signal is taken from LPAIR.  The background shape is taken from the
data in the left-hand plot.
\label{fig:ggpt}}
\end{figure}

A preliminary measurement of the cross-section for
muon pairs produced through two-photon fusion has been made 
by the LHCb collaboration~\cite{lhcbconf}
using the small 2010 data sample of $37 {\rm\ pb}^{-1}$.
The selection is as described in Sec.~\ref{sec:photpom} and the candidate events are
those in Fig.~\ref{fig:mu2mass},
with masses above 2.5 GeV but outside mass windows around the vector meson
resonances.
To determine the elastic CEP component, a fit to the transverse momentum distribution is made,
using a template shape from the LPAIR simulation~\cite{lpair,lpair1} to describe the elastic signal events
and using data, (low multiplicity dimuon candidates that have additional tracks,) 
to describe the background.
The  advantage of this approach is that it does not rely on the simulation of the background
processes that have a large uncertainty~\cite{lumi}.
The disadvantage however, is that it uses events in which the additional activity is seen
inside the detector (i.e. they have a relatively large $p_T$) 
and this probably does not have exactly the same $p_T$
spectrum as when the additional particles are produced very close to the beamline.
Nonetheless, a comparison of this data-driven background estimate to the simulation
of inelastic dimuon production, where one or both protons dissociate, 
shows good agreement (see the left plot in Fig.~\ref{fig:ggpt},)
albeit with rather large uncertainties
due to the limited statistics.
The fit to the signal candidates in the right plot of Fig.~\ref{fig:ggpt} also
shows good agreement and an almost
pure sample of dimuons from elastic diphoton fusion is obtained 
when requiring the $p_T$ of the pair to be below 100 MeV.
A cross-section times branching fraction estimate of $67\pm 19$ pb for both muons produced inside the LHCb
acceptance is in agreement with the theoretical prediction of \mbox{42 pb}, but is a long way from 
the aim of a few-percent measurement.

\begin{figure}[b]
\centerline{\includegraphics[width=0.49\linewidth]{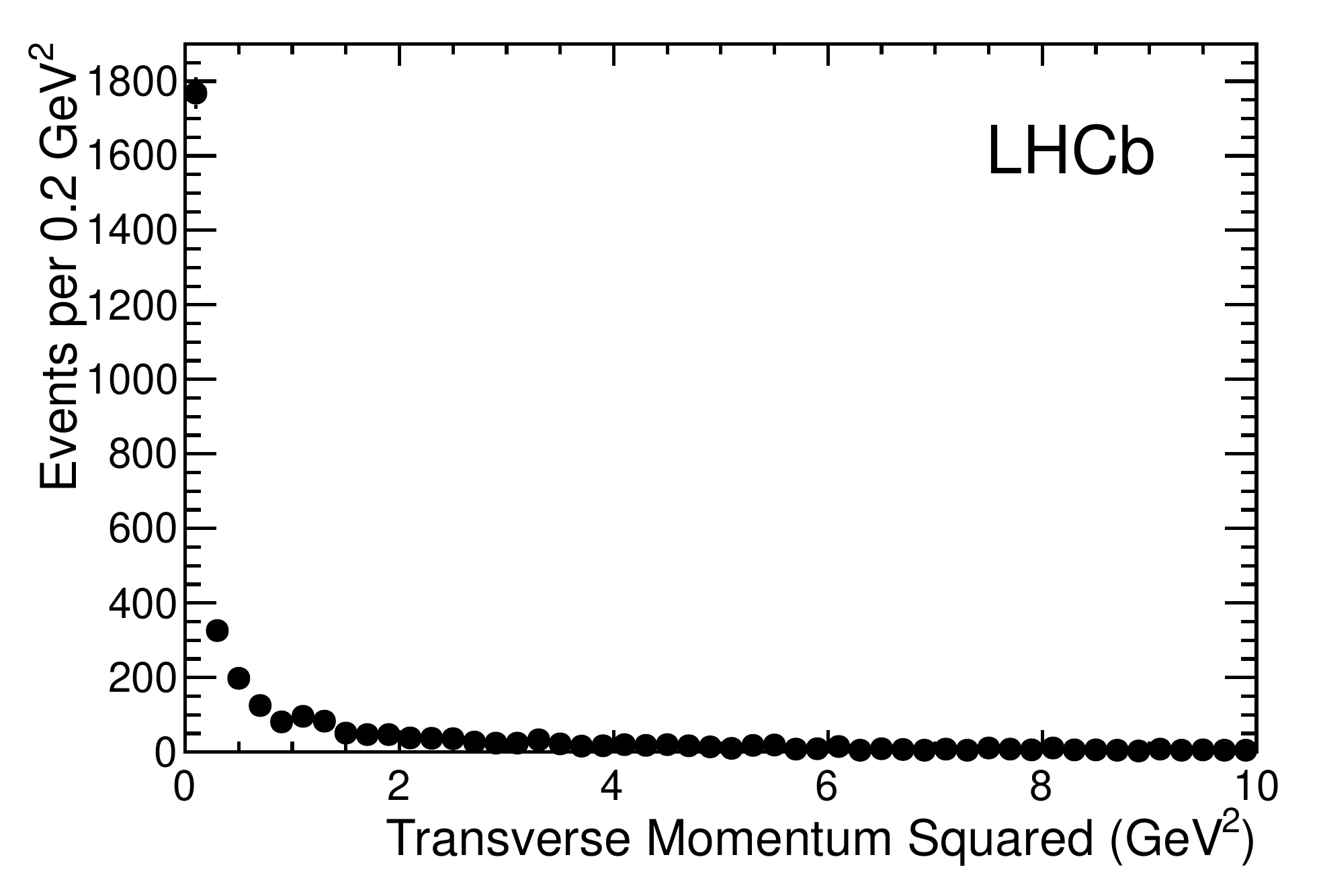}}
\caption{
Transverse momentum squared of dimuons 
with an invariant mass between 6 and 9 GeV.
\label{fig:mu2pt}}
\end{figure}

\subsection{Future analyses}
There are only 40 candidates with $p_T<100{\rm \ MeV}$ in the analysis of $37 {\rm\ pb^{-1}}$ of data, but improvements
to the trigger and a reduction in the average number of proton-proton interactions per beam
crossing suggest that about 10,000 candidates are available in the 
$3 {\rm\ fb^{-1}}$ of data taken in 2011 and 2012, sufficient for a 1\% statistical measurement.
Two things now become critical: firstly, whether the purity of the selected candidates can
also be known to 1\%; and secondly, what the precision on the theoretical value
for the QED elastic process is.
The former needs a more sophisticated approach than merely taking dimuon events with
additional activity to describe the background.
Possibly the theoretical shapes provided by LPAIR are sufficient to describe the inelastic
background, but this supposes that the only contributing processes are those described
in the generator as `semi-inelastic' and `fully inelastic'.
A hybrid approach using both theory and data is likely to provide the best solution
and there is hope that the required precision can be achieved.
Fig.~\ref{fig:mu2pt} shows the $p_T^2$ distribution for 
dimuon candidates with an invariant mass between 6 and 9 GeV in about
$3 {\rm\ fb^{-1}}$ of data taken at $\sqrt{s}=7$ and 8 TeV.
The strong peak below $0.2 {\rm \ GeV}^2$ is characteristic of the QED process.
Note that the distribution falls off much more rapidly than the $\sim \exp(-(6{\rm\ GeV}^{-2})p_T^2)$
dependence for the $J/\psi$ in Fig.~\ref{fig:pt2}.
An estimate of how much signal there is in the first bin requires
a complete description of the spectrum.

\section{Double Pomeron Exchange}
\label{sec:dpe}

DPE results in a hadronic central system that is a colour singlet and
has positive $C$ and $P$ parity.
For a central system with low transverse momentum 
this favours the production of a single meson~\cite{kmr_dpe,yuan}  
$q\bar{q}$ with
$J_z^{PC}=0^{++}$
or a pair of flavour-conjugate mesons~\cite{dicep}, 
$q\bar{q^\prime},q^\prime\bar{q}$.

\subsection{Single meson DPE}
Single mesons that should be produced by DPE include
$f_0(980),\chi_{c0},\chi_{b0}$ and their detection depends on their decay modes
and the competing backgrounds.
A favourable decay mode of the $\chi_c$ meson is to $J/\psi\gamma$, where the
only significant experimental background is contamination from $\psi(2S)\rightarrow J/\psi \pi^0\pi^0$
where only one photon is identified from the subsequent pion decays.
This is an attractive meson to search for as its mass is sufficiently high that its
cross-section can be theoretically predicted.

\begin{figure}[b]
\centerline{\includegraphics[width=0.49\linewidth]{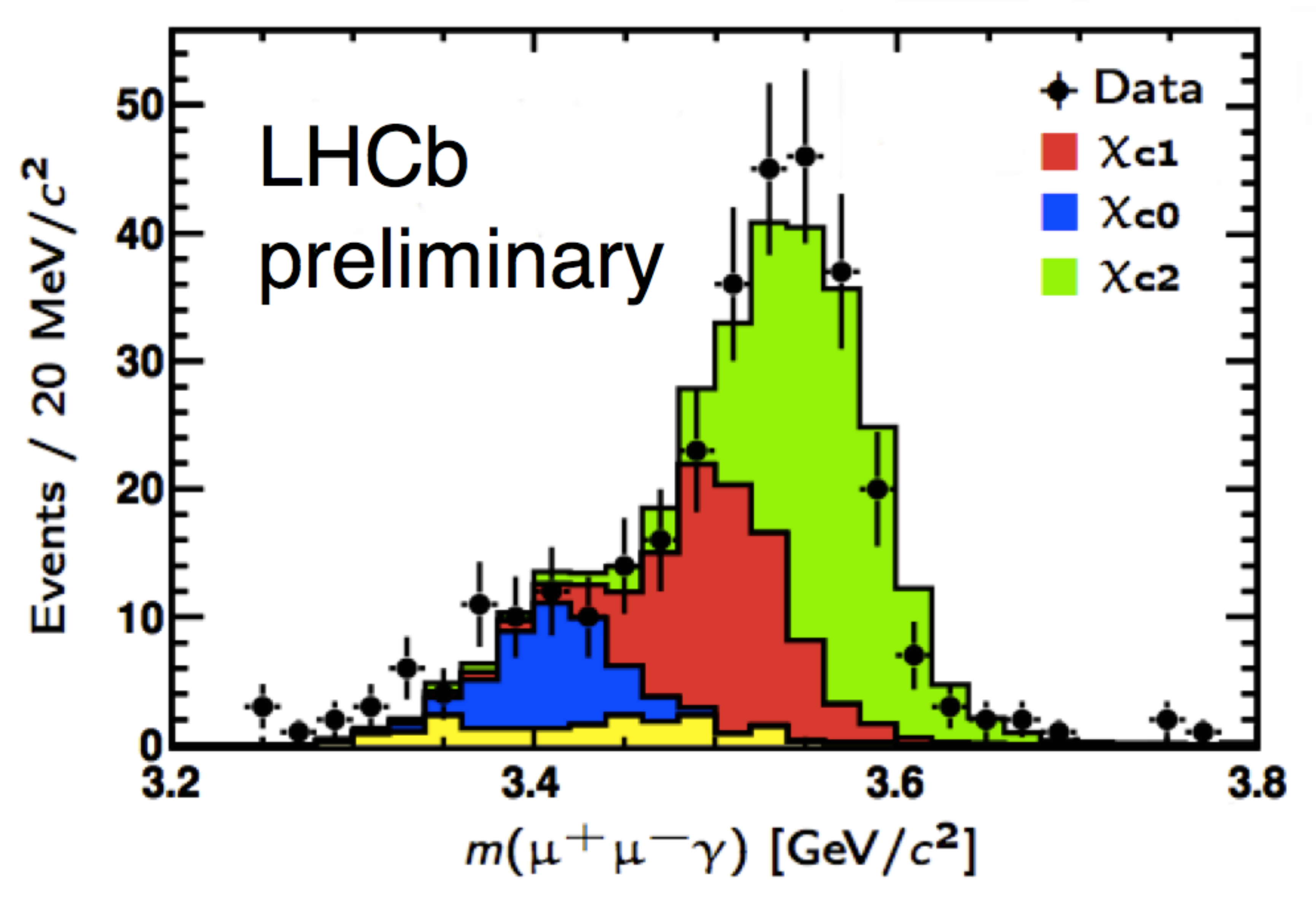}}
\caption{
Invariant mass of the dimuon plus photon system in events having no other activity
inside LHCb.
\label{fig:chic}}
\end{figure}

LHCb have made preliminary measurements~\cite{lhcbconf} of the
production of $\chi_c$ mesons with $37 {\rm\ pb^{-1}}$ of data.
The selection of events proceeds as for the $J/\psi$ selection in Sec.~\ref{sec:photpom}
but now one (rather than no) photon candidate is required.  The invariant mass of the dimuon plus photon system
is shown in Fig.~\ref{fig:chic}  fitted to expectations from the SuperChic simulation~\cite{superchic} for 
$\chi_{c0},\chi_{c1}.\chi_{c2}$ signal contributions and the $\psi(2S)$ background.
The CDF experiment made the first observation~\cite{cdf} of CEP of $\chi_c$ mesons but because
of the mass resolution, assumed it all to consist of $\chi_{c0}$ mesons.
The mass resolution of LHCb is sufficiently good to distinguish the three states.
In this decay mode, the contribution from
$\chi_{c2}$ dominates although much of that is due to the higher branching fraction
for this state to decay to $J/\psi\gamma$.
Unfortunately, the resolution is not
good enough to separate the three states completely and so
the fraction of the sample that is exclusively produced is determined for the
whole sample and is estimated to be
$0.39\pm0.13$ using the $p_T$ of the reconstructed meson.
The cross-sections times branching fractions are measured to be
\mbox{$9\pm5,16\pm9, 28\pm 12$\ pb }for $\chi_{c0},\chi_{c1},\chi_{c2}$, respectively,
slightly higher but in reasonable agreement with the 
theoretical predictions of 4, 10, 3 pb~\cite{superchic}.  
Only the relative cross-sections for $\chi_{c2}$ to $\chi_{c0}$ of $3\pm 1$ looks slightly 
higher
in the data than the theory expectation that they are roughly equal.
One possible reason for this discrepancy is that the fraction of elastic exclusive events in the sample
differs for each of the three resonances.
With greater statistics, a more sophisticated fit can be performed in order to estimate the
fraction of exclusive events separately for each $\chi_c$ state.

\subsection{Double meson DPE}

The production of light pseudoscalar and vector meson pairs 
has been considered
in Ref.\,[\citen{dicep}].
When the invariant mass of the central system is sufficiently high, a perturbative calculation
can be performed.
Exclusive meson pair production
is an interesting measurement in its own right in terms of understanding DPE
and the r\^ole of the pomeron, but can also shed light on hadroproduction and
meson wavefunctions.  Furthermore, it constitutes a background process for the
observation of $\chi_c$ decays to
pions or kaons as well as searches for glueballs or tetraquark states.

A theoretically interesting and experimentally accessible measurement is that of the
central exclusive production of pairs of charmonia.
The mass of the central system is sufficiently high that a perturbative prediction is
possible down to the threshold for production.
LHCb has recently made measurements of double charmonia~\cite{lhcblatest}, 
$J/\psi J/\psi, J/\psi\psi(2S), \psi(2S)\psi(2S), \chi_{c0}\chi_{c0},
\chi_{c1}\chi_{c1}$ and $\chi_{c2}\chi_{c2}$, using a data sample corresponding to $3{\rm\ fb}^{-1}$.

\begin{figure}[b]
{\includegraphics[width=0.49\linewidth]{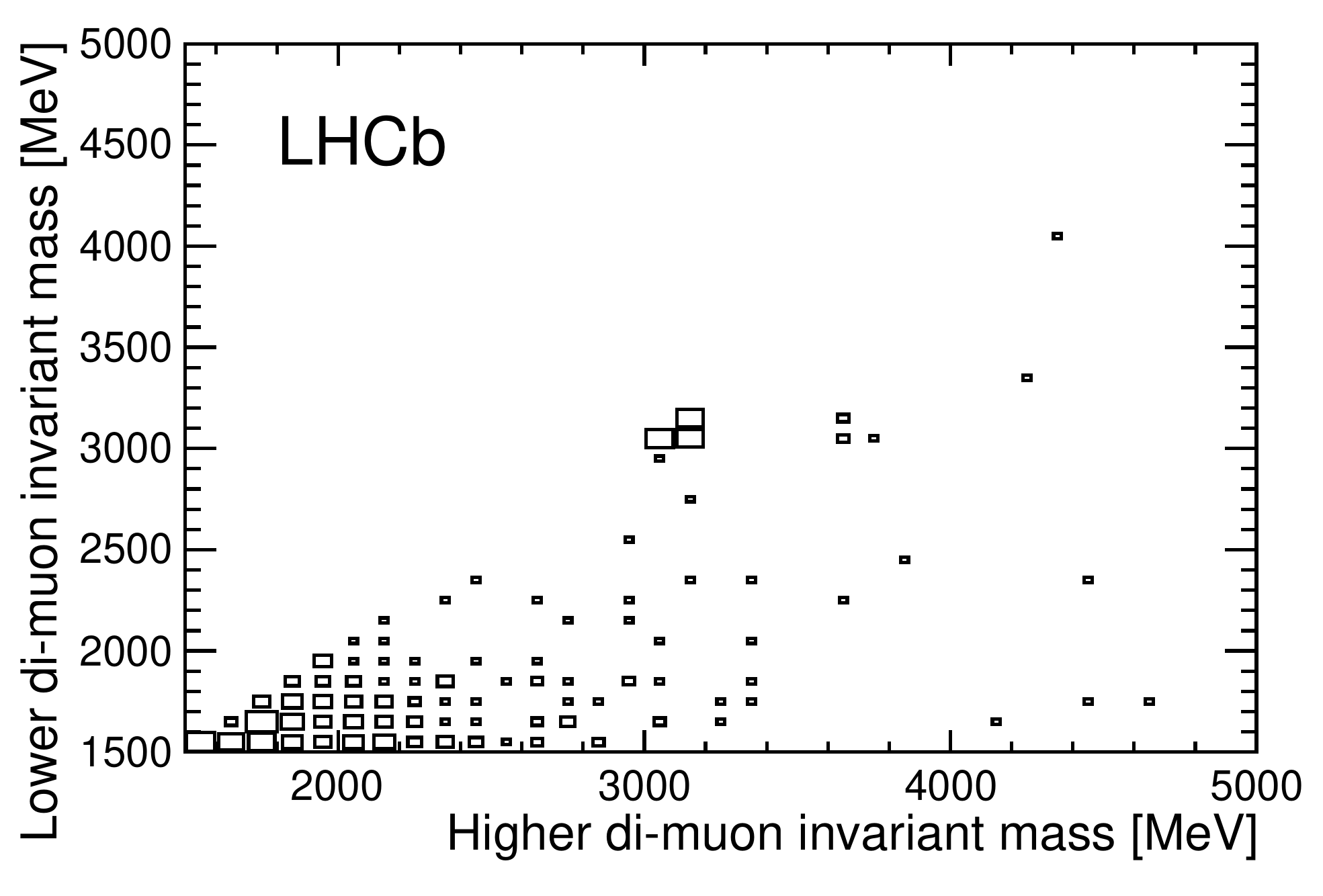}}
{\includegraphics[width=0.49\linewidth]{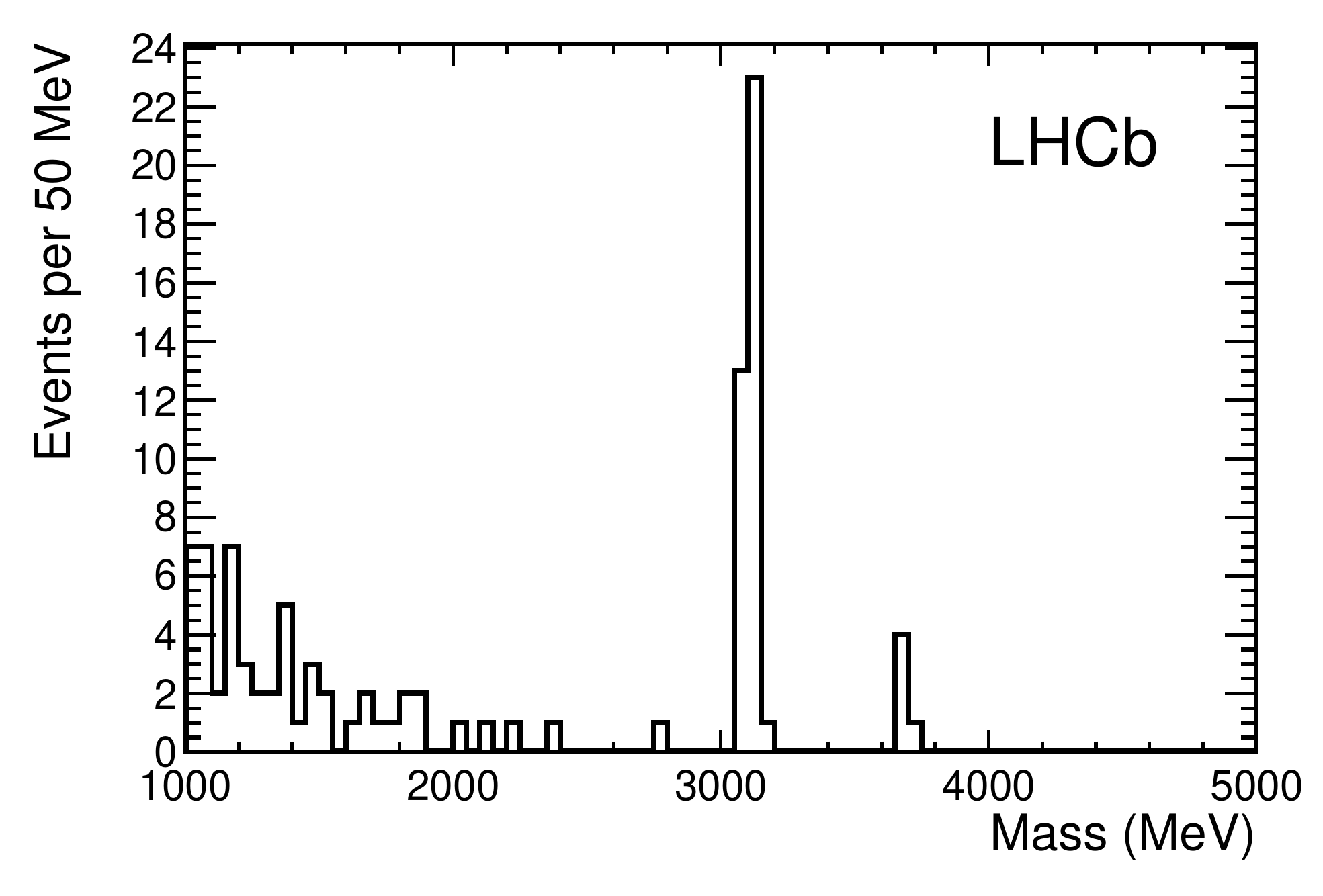}}
\caption{
(Left) Invariant masses of the two dimuon candidates.
(Right) The higher mass dimuon candidate having required the lower mass candidate to
be consistent with the $J/\psi$ mass.
\label{fig:2dmu}}
\end{figure}

The selection proceeds in a similar fashion to that described in Sec.~\ref{sec:photpom},
although now four charged tracks (at least three of which are identified muons) and no other activity are required to select pairs of S-wave states, 
while one or more photons are required to select pairs of P-wave states.
Very few low multiplicity events have three or more identified muons.  The invariant mass
distribution of the two pairwise combinations is shown in the left plot of 
Fig.~\ref{fig:2dmu} and shows
an accumulation of events at the $J/\psi$ and $\psi(2S)$ masses in a region of phase space that is
otherwise empty.  
The right plot in Fig.~\ref{fig:2dmu} shows the higher mass combination
when asking that the lower mass combination is consistent with the $J/\psi$ meson.
There are 37 $J/\psi J/\psi$,
5 $J/\psi\psi(2S)$ and no $\psi(2S)\psi(2S)$ candidates.
The only substantial background to the $J/\psi J/\psi$ signal comes from 
$J/\psi\psi(2S)$ where $\psi(2S)\rightarrow J/\psi X$ with $X$ unreconstructed.
After correcting for detector acceptance and efficiencies, 
the measured cross-sections for pairs of S-wave mesons with $2<y<4.5$,
which are exclusive {\it within the LHC acceptance},
are
$
\sigma^{J/\psi J/\psi} = 58\pm 10 \pm 6 {\rm\ pb} , 
\sigma^{J\psi\psi(2S)} = 63 ^{+27}_{-18} \pm 10 {\rm \ pb} $, and
$\sigma^{\psi(2S)\psi(2S)} < 237 {\rm\ pb}$ at the 90\% confidence level.
The search for P-wave pairs has a single candidate for $\chi_{c0}\chi_{c0}$ 
that is also consistent with
$J/\psi\psi(2S)$ production, and so upper limits at the 90\% confidence level
are set on the production of 
$\chi_{c0},\chi_{c1},\chi_{c2}$ pairs at 69,000, 45, 141 pb, respectively.

\begin{figure}[b]
{\includegraphics[width=0.49\linewidth]{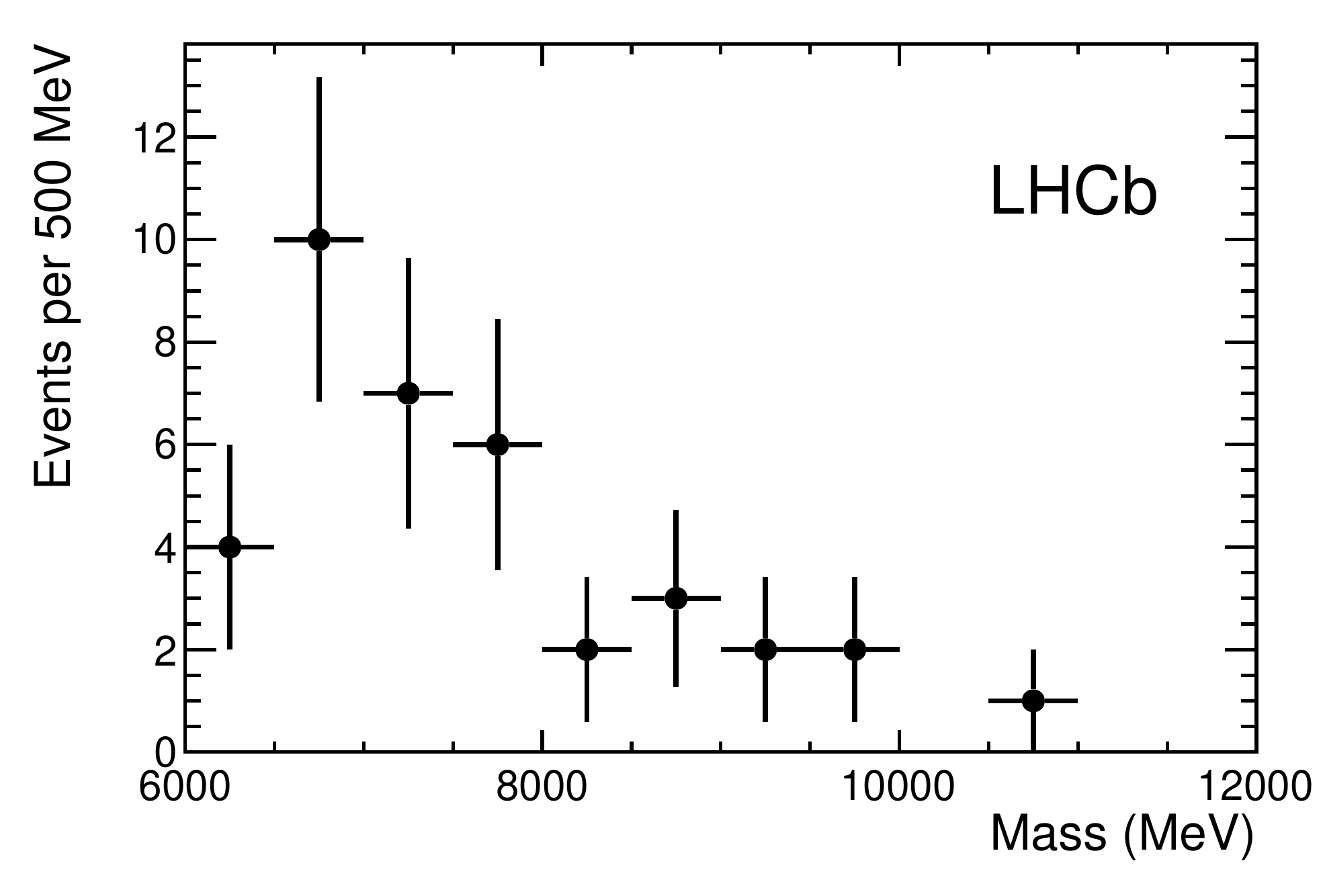}}
{\includegraphics[width=0.49\linewidth]{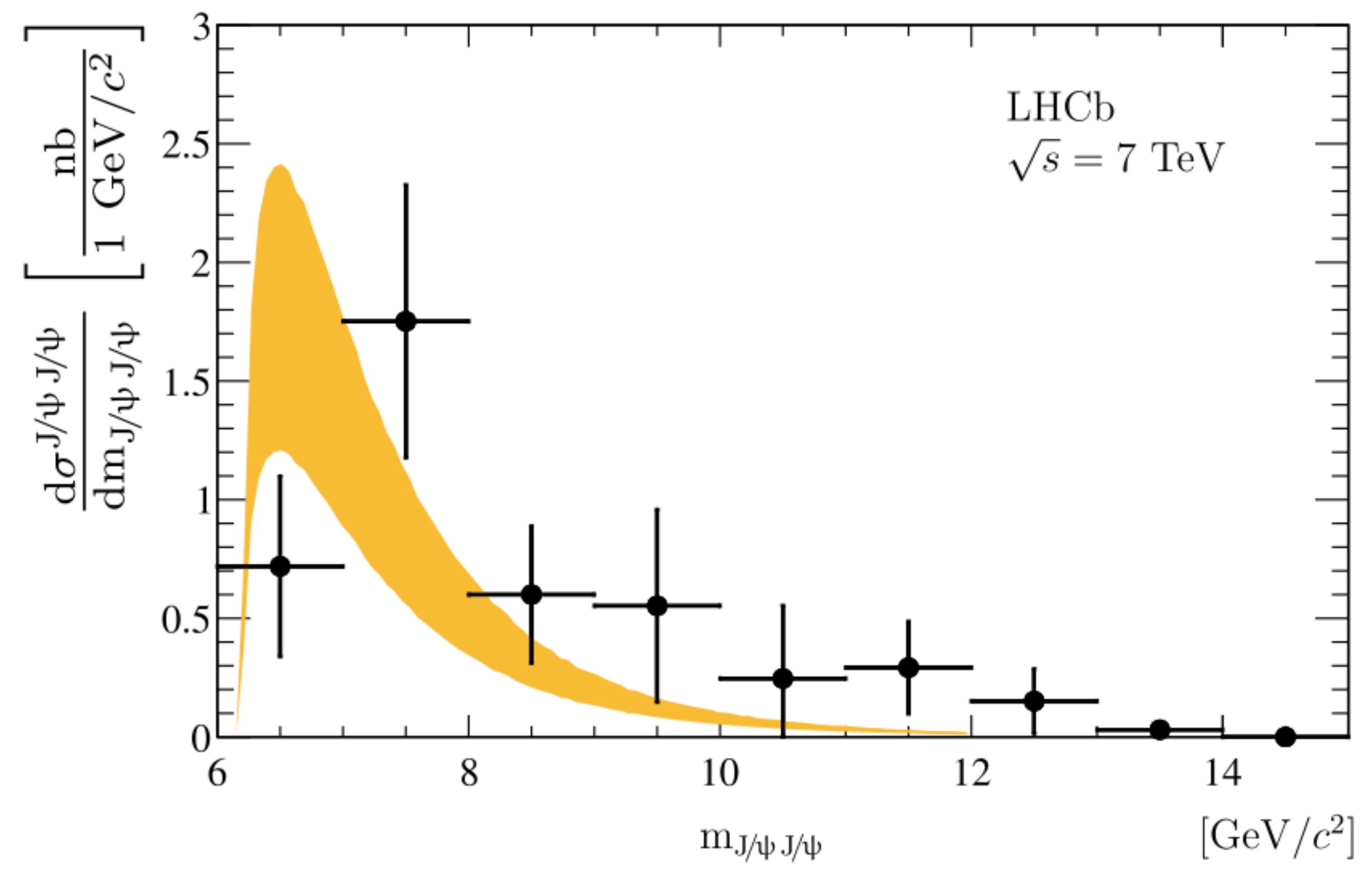}}
\caption{
Invariant mass of the $J/\psi J/\psi$ system in (left) exclusive
and (right) inclusive events.  The shaded area is the theoretical 
prediction of Ref.\,[\citen{berez}]
\label{fig:cfdjj}}
\end{figure}

The numbers quoted above are for dimesons detected in the absence of any other
activity inside the LHCb acceptance.  In order to compare with theory predictions,
a correction needs to be made for events that are not truly exclusive, as discussed in
Sec.~\ref{sec:exc}.  
This is determined
to be $(42\pm13)\%$ with a large uncertainty due to the low number of $J/\psi J/\psi$ events observed, and
leads to a measurement of elastic CEP $J/\psi J/\psi$ with $2<y<4.5$,
at an average $\sqrt{s}=7.6$ TeV,
of $24\pm 9$\ pb.
This is in agreement with 
a preliminary calculation of 8\ pb that has been obtained~\cite{lucian}
following the approach of \mbox{Ref.\,[\citen{dicep}].} 
There is a large uncertainty on the theoretical number coming from the poorly
understood low-$x$ gluon PDF that enters with the fourth power in the theoretical calculation.
More data, both to pin down the
gluon PDF (as described in Sec.~\ref{sec:photpom}) and to improve the $J/\psi J/\psi$ CEP measurement, will enable a more precise 
comparison.  

In Fig.~\ref{fig:cfdjj},
the invariant mass of the exclusive signal is compared to that of an inclusive measurement
of double $J/\psi$ production, performed by LHCb~\cite{lhcbdoublej}; both have a similar shape.
The data in the inclusive measurement are shifted to slightly higher masses than the theory,
and this has been discussed as possible evidence for
double parton scattering~\cite{stirling}
or tetraquark states~\cite{berez}.  The former is negligible in DPE due to the ultra-peripheral
nature of the collision, and thus with more statistics, the exclusive measurement 
will become sensitive to the presence of
higher mass resonances.

\subsection{Future analyses}
With almost 100 times more data collected than that used in the 
preliminary $\chi_c$ analysis,
several additional measurements can be performed to determine the relative production of each
of the $\chi_c$ states.
Firstly, as mentioned above, the analysis can be repeated with greater statistics allowing
the exclusive fraction of each $\chi_c$ state to be determined separately.
Secondly, the resolution of each of the states can be improved by reconstructing photons that
converted into electron-positron pairs in the detector material.  This proved very successful
in the LHCb inclusive analysis of $\chi_c$ states~\cite{lhcb_chic}.
Thirdly, the $\chi_c$ can be reconstructed in different decay modes.
Of particular interest are the decays to two pions or two kaons, which are not possible for $\chi_{c1}$
and are about four times higher
for $\chi_{c0}$  than for $\chi_{c2}$.
In addition, the mass resolution in this channel is about a factor of three better than in the
$\mu\mu\gamma$ channel.
The LHCb preliminary or published CEP analyses to date have concentrated on
final states with dimuons because these are very clean.  
However, a large amount of low-multiplicity data has been collected, triggering on electromagnetic
and hadronic energy.
Consequently, the observation of $\chi_c$ states in the $\pi\pi$ and $KK$ modes ought to
be possible, so long as the backgrounds from the DPE production of pairs of pseudo-scalar 
mesons is not too large~\cite{dicep}.

Following on from the observation of pairs of charmonia, measurements should also be possible
of pairs of $D$ mesons.  The software trigger explicitly selects $D$ meson decays to pions
and kaons in a low multiplicity environment.
Light meson spectroscopy is another topic where measurements from LHCb can be expected.
Pairs of pions and kaons are triggered if the summed transverse energy is higher than $\sim 1$ GeV.
For low transverse momentum objects, this allows the dipion and dikaon mass spectra
to be reconstructed for masses above about 1 GeV.  This region is rich in resonance structures
as observed at the ISR~\cite{isr} and Tevatron~\cite{mike}.  The excellent mass resolution and particle identification
of the LHCb detector mean these should be easy to see, although the interpretation of
the overlapping structures requires a dedicated programme of work.

\section{Summary and Outlook}
\label{sec:summary}

\begin{figure}[b]
\centerline{\includegraphics[width=1\linewidth]{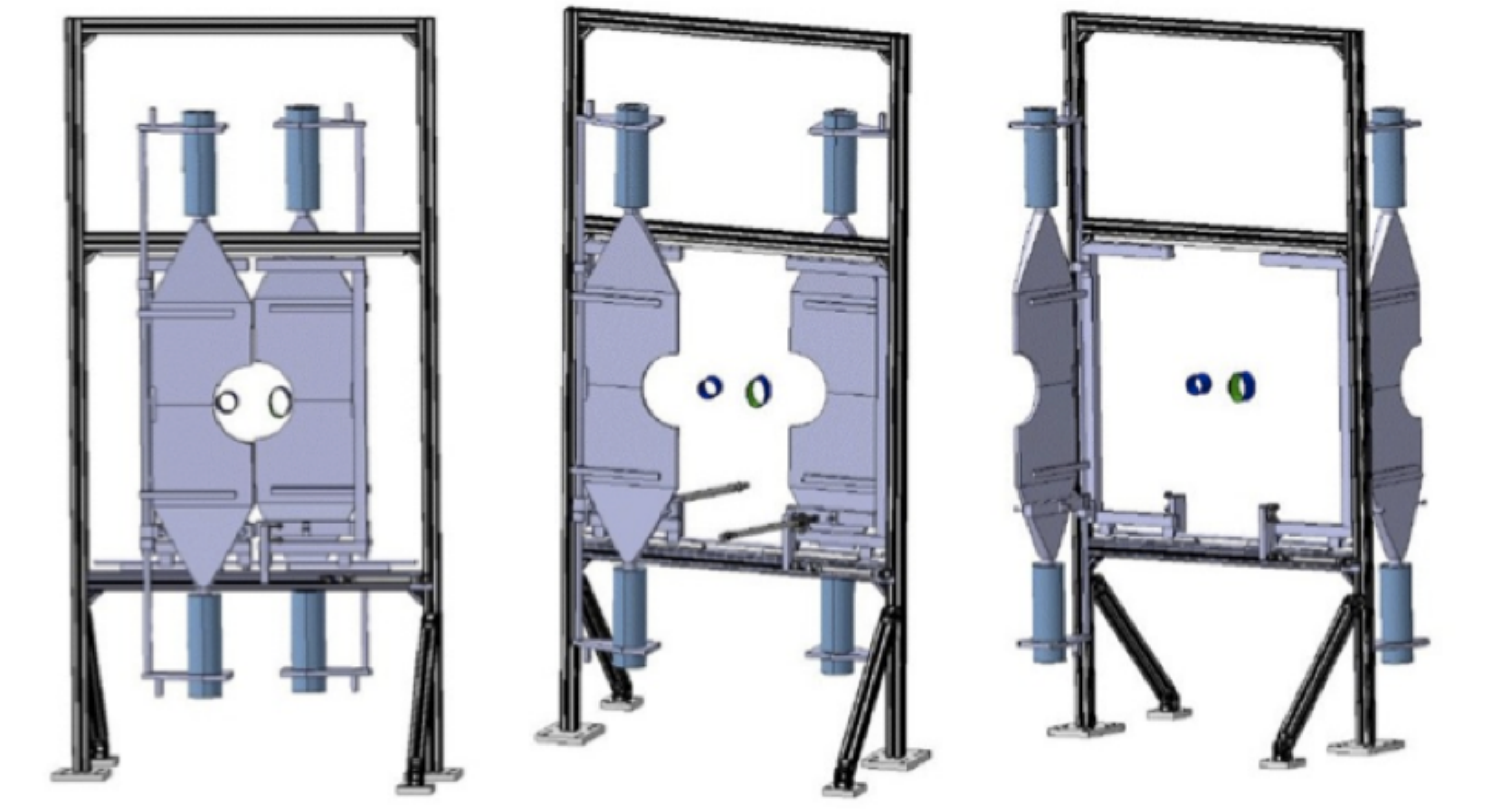}}
\caption{
Drawings of one plane of the Forward Shower Counters
are currently being installed 
around the beampipe,
in the LHC tunnel, on either side of the interaction point.
The picture shows the support structure, scintillators and PMTs, which can be retracted
from the beamline in case of adverse conditions.
\label{fig:herschel}}
\end{figure}

The design of the LHCb detector for b-physics makes it well suited to the exploration of
CEP.  It has excellent particle identification and 
the ability to trigger on low transverse momentum objects.
The coverage of the detector is reasonably good but is proving the limiting factor in making
measurements of CEP.  In particular, there is no coverage above $|\eta|=5$, which is precisely
the region that particles originating from proton dissociation are likely to occupy.
Consequently, new Forward Shower Counters (FSC)
are currently being installed to take data when the LHC
restarts in 2015.  They consist of several layers of plastic scintillator placed on either side
of the interaction point at distances of 7.5\ m, 19\ m and 114\ m.  
A schematic of one station is shown in Fig.~\ref{fig:herschel}, in which can be seen the hole 
through which the beampipe will pass,
the scintillators, and PMTs.  The detector has been designed to be retractable
when beam conditions deteriorate.
Geometrically this new detector extends
the pseudorapidity reach out to 8 units, but simulation shows its impact is even greater
as the showering of particles at higher pseudorapidities also leaves signals in the scintillators.

Information from this new detector will greatly improve LHCb's 
ability to distinguish elastic CEP events 
from those with proton dissociation.
There is also the possibility of including the output directly in the trigger and
vetoing on activity in the FSC, thus increasing the bandwidth available to collect CEP events.

CEP was not originally considered in the design of the LHCb detector.  
However, with the inclusion of some simple triggers to select low charged multiplicity events,
many CEP measurements have become possible.  The published measurements to date,
including the first observation of pairs of charmonia, are the start of a broad range of 
measurements currently underway with both leptonic and hadronic final states.

\section*{Acknowledgements}
The author wishes to acknowledge the support of Science Foundation Ireland.

\end{document}